# Handling Large and Complex Data in a Photovoltaic Research Institution Using a Custom Laboratory Information Management System

Robert R. White[1] and Kristin Munch[1]

[1]National Renewable Energy Laboratory, 1617 Cole Blvd., Golden, CO 80401, USA

## ABSTRACT

Twenty-five years ago the desktop computer started becoming ubiquitous in the scientific lab. Researchers were delighted with its ability to both control instrumentation and acquire data on a single system, but they were not completely satisfied. There were often gaps in knowledge that they thought might be gained if they just had more data and they could get the data faster. Computer technology has evolved in keeping with Moore's Law meeting those desires; however those improvements have of late become both a boon and bane for researchers. Computers are now capable of producing high speed data streams containing terabytes of information; capabilities that evolved faster than envisioned last century. Software to handle large scientific data sets has not kept up. How much information might be lost through accidental mismanagement or how many discoveries are missed through data overload are now vital questions. An important new task in most scientific disciplines involves developing methods to address those issues and to create the software that can handle large data sets with an eye towards scalability. This software must create archived, indexed, and searchable data from heterogeneous instrumentation for the implementation of a strong data-driven materials development strategy. At the National Center for Photovoltaics in the National Renewable Energy Laboratory, we began development a few years ago on a Laboratory Information Management System (LIMS) designed to handle lab-wide scientific data acquisition, management, processing and mining needs for physics and materials science data, and with a specific focus towards future scalability for new equipment or research focuses. We will present the decisions, processes, and problems we went through while building our LIMS system for materials research, its current operational state and our steps for future development.

## INTRODUCTION

The scope and capabilities of computers to support scientific research and experimentation has grown enormously in the past fifty years. These systems have now become integrated tightly within most experimental systems from controlling the instrument operations to data acquisition and in some cases analysis. While in general this has improved greatly our abilities to utilize computers to meet our needs in research; it has created a deluge of data caused by large scale resolution imagery, high resolution temporal data, extremely large or high dimension data sets, and advanced simulations and modeling. Data streams can now deliver hundreds of megabytes if not gigabytes of data very quickly, but our ability to effectively store and process the data has not kept up. This is not only an issue in photovoltaics but has touched across many fields in the sciences [1-4]. In addition distributed research across our lab and institutions highlights the need to efficiently share the resources and data [2]. Preparing cloud ready data streams is becoming an important process at many companies and academic and government institutions [2, 3]. Losses due to many factors including corporate cultures, system failures, lack of reporting negative results and a lack of reporting of full data sets are ongoing issues of data handling occurring not only in our institution but others too [5]. All of these losses impinge on our ability to gain a complete picture from the data.

To tackle the data processing issue, many are using data mining to provide effective means to analyze the data through a variety of algorithms and artificial intelligence from support vector machines through decision trees to Bayes classifiers. Yet the quality of the products produced by the analytics is only as good as the quality and completeness of the data feeding them. Being able to produce, house, aggregate and transfer the data properly will enhance the work done through analytics and this infrastructure goal has now become a major issue for all big data handling within the scientific community and without.

The National Renewable Energy Laboratory (NREL) and the Department of Energy have begun a focus on handling big data, not only looking at analysis but how to handle the raw data effectively and to create databases of information that are efficient enough to enable science and education into the next century. This presents several challenges due to the size and mutability of the lab environments at NREL. Our labs are dynamic environments where instruments are added and removed on a regular basis. Any data handling system must adapt to new resources placed online with a minimum of code rework. Most instruments are designed by manufacturers as standalone systems with minimal integration capability, often limiting their capacity to produce a quality data stream. Strict cyber security guidelines also pose issues with how and where data is stored and accessed. Some of our older instruments have antiquated operating systems presenting challenges in network integration; where they must be isolated from the greater institution network. Further confounding the problems is that there are no data format standards for photovoltaic research equipment where instruments with the same capability often have radically different methods to report data.

**APPROACH**

In order to efficiently handle large heterogeneous data products for the solar program, we needed a LIMS that could handle all of these factors discussed. We also had to objectively weigh the amount of time and resources needed to design and implement this software or find a consumer solution that could agilely adapt to our laboratory's current and future needs.

We performed an initial evaluation of several available LIMS. While many offer very compelling features and capabilities, they all failed to meet our needs in two important categories. The first was that off-the-shelf LIMS are typically focused on one particular industry and to fit our needs would require extensive customization. Secondly with our instrumentation changing on a regular basis, any tables in the supporting database and any user interface would need to be updated periodically. Agile customization of an existing deployment would be expensive and often not timely. The added extensive customization required would reduce any advantage from a ready built solution.

But there are some advantages to the off-the-shelf LIMS solution: lower overall costs, feature rich environments, and maintenance contracts, yet these were often outweighed by the lack of target features for our specific areas of science. We made the decision that to achieve our goals we would need to design and construct a LIMS in-house, doing the code development as a joint project between the National Center for Photovoltaics (NCPV) and the Computational Sciences Center (CSC) at NREL. With the domain expertise from the NCPV, data modeling for the instrumentation could be handled quickly. The CSC would not only provide development expertise but would also be an end-user; helping researchers with analytics for data mining and visualization.

**Design**

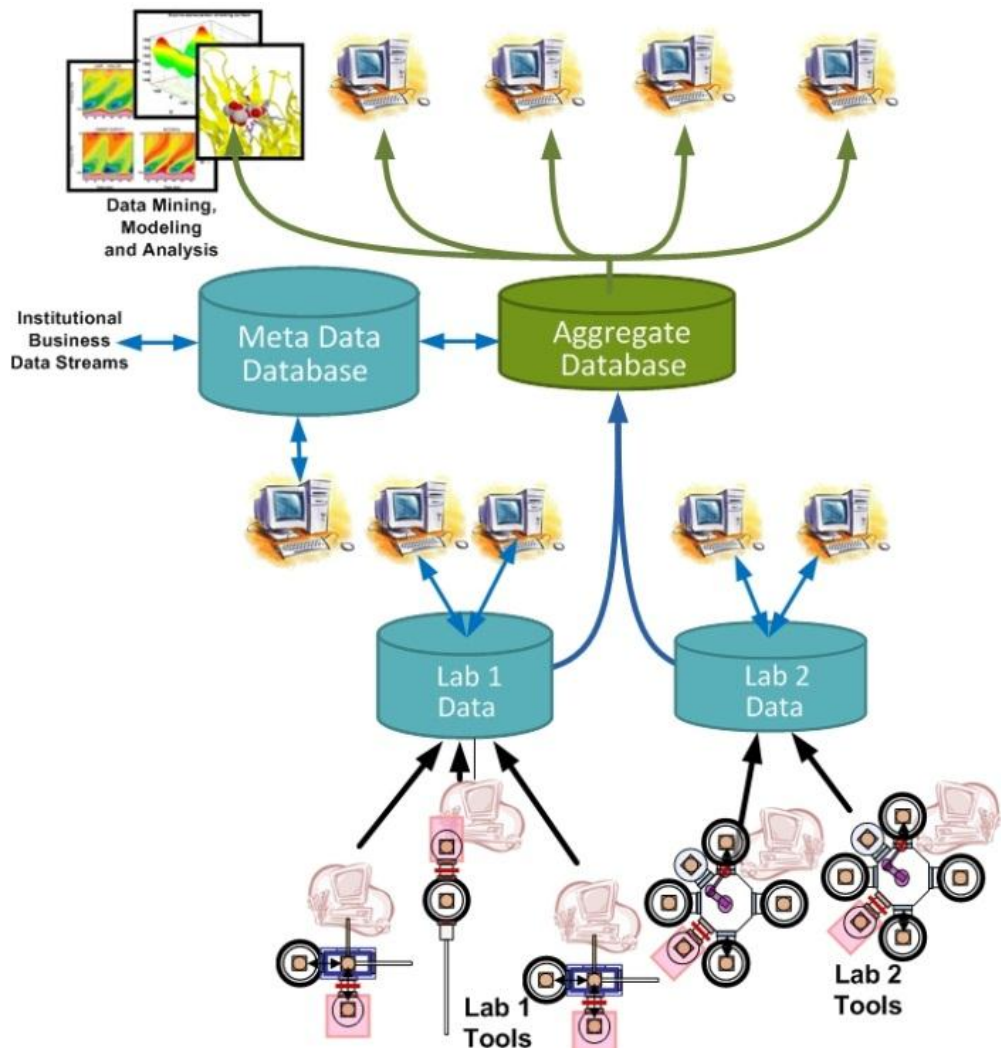

**Figure 1 -** Hierarchical structure of a modular LIMS system

The LIMS that we wished to design was based on the layout in Figure 1. We chose a layered approach where each lab operated independently, including harvesting and storage of their data resources. In time each individual lab system will be merged under a master application to provide a single point to access and mine all data products. Even with the master access point, individual labs will still be able to directly access their particular database and archive. Instrument computers from each lab are independently monitored and harvested to a local centralized server. There the data is stored in a file archive by tool and date; providing a course informatics granularity. Harvested data is then further extracted from the files and aggregated into a relational database. The entirety of the data is presented via a web interface formatted into concise and easily searchable pages. The database tables are also accessible directly through SQL queries sent from analytical software packages. Through that same web interface, researchers are also able to annotate and associate the raw data with metadata for tracking experimental information, sample details, and publications.

The design will eventually grow to provide all solar program labs with informatics support. Due to the size of the solar program and its instrument portfolio this would be a daunting task for a small number of developers. However, the layered design approach allows labs to be brought fully into the informatics system individually, one at a time. Issues, bugs, and lessons learned can be addressed during each subsequent lab start up, making the process of integration quicker and smoother over time. An additional bonus to the layered design is that failure of any particular lab hardware or software would not precipitate data issues in the overall LIMS and any faults can be cleared up locally and lost data post-processed into the system.

Owing to the manner in which some of our processing and characterization tools operate it is necessary to choose a means for identifying and harvesting files that is unobtrusive and unlikely to cause any sort of fault or failure of the tools. Processing systems that deal with volatile gases and complex procedures cannot be interrupted without creating possible safety hazards. Characterization tools can run for extremely long periods and interruptions mean a loss of time and resources. Simple programs running as remote processes perform the monitoring of the instruments. We chose to have a centralized file harvesting program controlling data flow between lab instruments and the server to avoid interruption of the local data capture on instruments. Data translation and database storage would be the most computer intensive activity, so having a system that could operate independently of the harvesting and monitoring would be the most practical. We also identified several small project oriented databases that had been fielded over the years. Merging these into a new databases or effectively linking the existing tables of this highly compartmentalized data would be required to complete an overall informatics system.

From our design requirements we saw the need to build the system in modular blocks. This provided additional flexibility by allowing us to change or improve sections of the code efficiently without disrupting the rest of the software base. Creating independent software blocks meant we needed to design a communication format and decide on a transfer protocol. It was also possible that some of these blocks would be deployed on different computer systems, so any messaging would also need to be able to communicate across the network. We chose XML as our communication format due to its wide spread adoption, ANSI standardization, and the fact that nearly every programming language has support for XML encoding and decoding. We developed two standard XML schemas to support our operations: an operation message to handshake between the software modules and another to provide a common data format for translated instrument data. For a communication protocol we use basic TCP/IP due to its simple integration and robust error handling.

## **Implementation**

The back-end software code that performs the harvesting and data extraction is deployed on a Linux Operating System (OS) configured as a LAPP (Linux + Apache + PostgreSQL + PHP) stack deployed within the lab. The PostgreSQL database is deployed on a similar system within the CSC data center. The database deployed monitoring scripts communicate through the lab server via secure protocols to the monitored instrumentation.

Each monitored instrument has a hard drive share or shares CIFS mounted to the local lab server, which can then directly access the directory structures of the instrument. By not deploying software on every instrument, we remove the need to build and maintain several different versions of software for the variety of operating systems controlling the instruments. In addition, we remove the possibility of causing problems with that instrument's computer OS. We have seen issues of directly deployed software causing crashes or command and control software failures previously. Except for the access setup and preparing the remote share, the instrument computer is left in a pristine state.

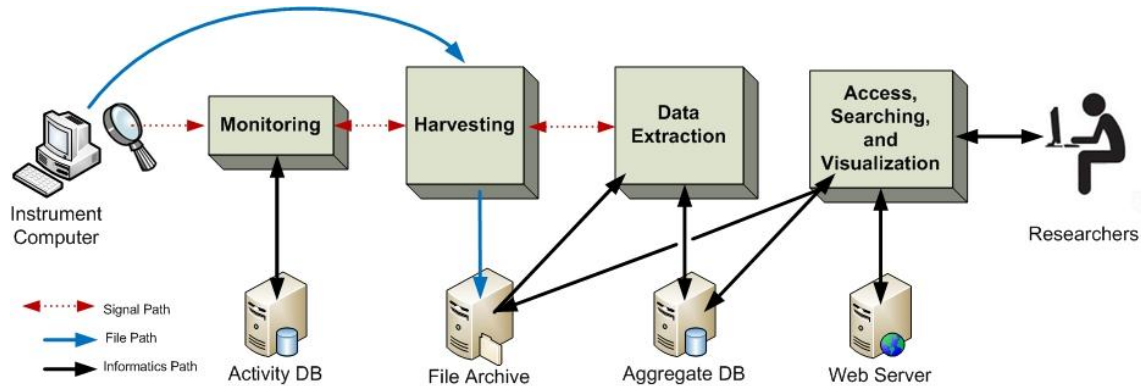

**Figure 2 -** Signal, File and Informatics path from the instrument to the researcher.

    The code architecture is composed of modular blocks, as designed. The primary part of the back end code consists of three software components: a monitoring program, a harvesting program and a data extraction program. The monitoring program identifies new files of interest, reports findings to the harvesting program and records transaction time and any errors reported back by the harvesting program. The monitoring program also keeps track of any changes in file size to indicate updates to any previously harvested files. An updated file is also harvested and then swapped with an existing file in the archive. The second component, the harvest program, copies files out of the tool and onto the lab server file system, creating a default secondary backup and applying a first layer of security or encryption if requested. The harvesting program also controls the priority of file transport based on defined priorities and current network traffic. The third component, the data extraction program, is alerted when a file is copied successfully by the harvesting program. Based on requirements defined in a configuration file, the harvested file will be opened, translated, and relevant data extracted into relational database tables.

    Once the files are archived and data is extracted, the information is available to the researchers through the front end web interface. The researcher is presented with a set of pages showing the tool, projects and samples, that they are currently authorized to see or use. We are also in the process of building the methods to provide an interface for users to be able to track and search new or existing projects, samples, tools, or data including a means to upload analysis data to associate with projects or samples.

    Each of the back-end software components and monitoring scripts communicates via XML messaging. All notifications and acknowledgements are passed through the bound TCP/IP sockets, with the TCP/IP server being established by the harvesting software and the monitoring and translation/data extraction software as clients. The messaging consists of a two packet burst; one containing number of bytes in message followed by the actual message. This helps with error handling in identifying incomplete or delayed messages. The harvesting and data extraction programs perform verification, once a message is received, against the XML schema.

```xml
<?xml version="1.0" encoding="UTF-8"?>
<NCPVOps role="transfer" timestamp="2001-12-17T09:30:47Z" xsi:noNamespaceSchemaLocation="NCPVOps_Schema.xsd" xmlns:xsi="http://www.w3.org/2001/XMLSchema-instance">
  <Target>
    <Source path="/mnt/bruker/frames/VCC_1234.txt" host="nexus"/>
    <ToolInfo name="XRD Bruker" type="Characterization"/>
    <SampleID>cigs-0013-00023</SampleID>
    <Operator>
      <Username>jdoe</Username>
    </Operator>
  </Target>
</NCPVOps>
```

**Figure 3** - Example of communication message between monitoring software and harvesting software identifying a new file to be copied to archive

## Common Data Format

The heterogeneous instrumentation within the solar program labs produces large amounts of data in widely varying formats, requiring translation into a common data format to facilitate efficient utilization within the LIMS. There are some similarities of data type, but the layout and structure varies across almost all instruments. Some instruments are custom built by our engineering staff and we have predefined those data products, but many of the remaining instruments are off-the-shelf systems, usually with proprietary and sometimes opaque data formats. To efficiently handle the heterogeneity of data formats across instruments, we designed a common data format for collecting and translating instrument data.

By studying the work flow for both a generic processing and characterization system, we noticed a similarity in the data elements that could lead us to a common data format. Evaluating instrument data formats at the most general level shows the input to any complex instrument to be a summation of an assortment of settings that control the activity. In the case of processing this would be a recipe containing a series of steps each encompassing controls for a set of hardware. For characterization equipment this same pattern can be seen as defining the instrument setup and control. At the end of a run the system produces a set of data from various diagnostic tools, some temporal some not. In the case of characterization systems the desired result is the set of measurements and in the case of processing; a series of diagnostics taken over the run. Using this idea we could now build a XML schema to match the generalities of any activity with elements to make specific identifications as needed.

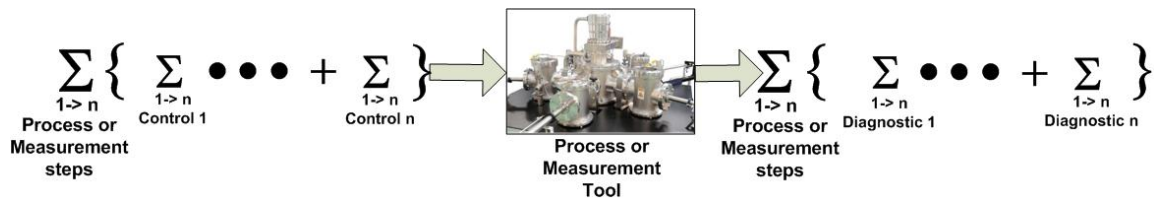

**Figure 4** - Commonality in the assembly of processing steps and characterization instrument setups. The results are a series of step defined diagnostic data from a single or multiple pieces of instrumentation.

But one additional issue that needed to be confronted when building the translator is that many off-the-shelf instruments lacked self-described data; a value within the data file easily being matched to the name or type of data and units. Some of these decisions may have been made from marketing strategies to discourage competition; nonetheless this tactic causes difficulties when creating an integrated data environment. It is not uncommon to find the data coming from some instruments to be a simple series of columns with no descriptions at all. In some of our cases we used a bit of deduction to determine what the values are describing. In others we were able to request the information from the manufacturer. Once we have the data format produced by the instrument and the schema for the common data format we could then build the universal translator.

**The Translation Process**

We have chosen XML as the language for the common data format in our LIMS. The XML schema is scalable and can encompass everything from the smallest amount of metadata associated with a transaction to all the data points in a large multi-dimensional data file. We constructed a program that could adapt its process of translation based on what instrument the file came from and the type of data. The translation parameters are kept in configuration files and identify the data structure and important elements. We defined the database tables to closely mimic the common data structure and the translation libraries provide simple methods for pushing the data from the XML file into the database. Data can also pass back through the translator from the database to produce a "hard" copy of the data. Some instruments produce binary formatted data products (.pdf, .img,, etc.) and we have opted to not extract information directly from those files. Instead the translator will collect as much meta-data as possible from the transaction and record this. As resources become available we may return to implement data extractors for these files where needed.

To optimize the performance of the LIMS we execute translation and data extraction immediately post-harvest. As designed, a separate software module running independently of the harvest and monitoring performs the data extraction, allowing harvesting to continue independently of extraction processes. Communication between the harvester and the data extractor is via a standard XML operations message. These messages trigger activities within the translator based on the role attribute contained in the message's root element. The data extractor can be commanded to translate a new file, update an old one, read data from the database and transmit it back, or any one of a series of test functions.

The XML element fields inherently carry information on where in the database the information is to be stored. This translator allows for agile scalability; tools added onto the LIMS can begin immediate harvesting and researchers can access to the raw data within a few hours. We have seen some lags due to large amounts of historical data that must be retrieved from a tool at first, but then the harvesting and extraction runs quickly.

```xml
<?xml version='1.0' encoding='UTF-8'?>
<NCPVData xmlns:xsi="http://www.w3.org/2001/XMLSchema-instance" role="archive" timestamp="2012-09-25T12:45:03"
id="" xsi:noNamespaceSchemaLocation="../PDILData_Schema_v40.xsd">
 <Characterization>
  <Type id="1" name="Reflectance and Transmission"/>
  <Tool id="23" name="N and K"/>
  <Operator id="7"/>
  <DataFileLink timestamp="2012-09-25T12:45:03" file="nandk/data/20130108/azo_azo_a239_output.1"/>
  <Comments></Comments>
  <Aggregate>
   <MetaData units="position" value="0.0000" name="x"/>
   <MetaData units="position" value="0.0000" name="y"/>
   <MetaData comments="RTnk" units="-" name="method"/>
   <Data>
    <descriptor units="nm" name="Wavelength">
     <datum value="1000.00"/>
     <datum value="999.00"/>
    </descriptor>
   </Data>
   <Data>
    <descriptor units="exp" name="Reflectance">
     <datum value="0.096500"/>
     <datum value="0.096100"/>
    </descriptor>
   </Data>
  </Aggregate>
 </Characterization>
</NCPVData>
```

**Figure 5 -** Example of a common data format file. Each transmission can contain multiple data blocks if needed, with each wrapped inside the "Aggregate" element. Within each Aggregate are a set of self-described data elements. There is no limit on the number of data points or sets of data points in each Aggregate element.

## **Database and Data Modeling**

The database was implemented with three major elements: a metadata structure to capture overarching information pertaining to all LIMS data, a generic structure for variable definitions and a semantic structure for more rigid definitions concerning certain instruments. The metadata structure of the database was built to take advantage of natural relationships that occur in PV research: samples, projects, processing procedures, instrumentation, measurement procedures and sample storage methods; information which can provide simple yet valuable insight into differences in the performance and processing of any cell. The physical database design encompasses elements from both generic and semantic logical data modeling efforts. We employed a combination of generic and sematic modeling to enable both initial flexibility of adding new instruments onto the system as well as the eventual known semantics of instruments to meet query and display needs. The generic data models enable scalable and variable instrument table definitions and allows for unknown and variable data types to be collected, parsed and stored in the database tables. The semantic data models provide relational data structures for known data formats and allow for instruments with standard output to be physically represented in the relational database using the instruments semantics as table and column definitions. The semantic representation of an instruments data provides the means for efficient query, plotting and mining of the data. Often tools move from being defined initially with a generic model to eventually being defined semantically to enable searching and plotting of data through the use of dashboards and interactive displays.

In the generic structure of the database, we looked at commonality in events of a sample's lifetime (processing and measurement) and adopted a common tactic used in some biological relational databases and LIMS, by making use of generic data modeling techniques. We constructed tables which scale according to an instrument's data in order to accommodate a variable number of fields and data points from activity to activity. In addition, the generic structures take advantage of object-oriented inheritance capabilities of the relational database. We created base type tables with common characteristics and specialized tables that would inherit the base elements and expand on those to match the targeted instrument. This aided greatly in expediently integrating instruments that are closely related in capability and function.

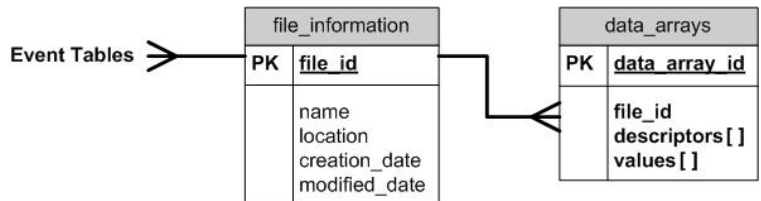

**Figure 6** -The generic data model allows for flexible data table design. In this example, we take advantage of PostgreSQL database's ability to handle arrays as part of the table data types. Anytime a file is added to the database its basic meta-data is stored in the file information table and linked back upstream to the actual event (processing or measurement) that created it. From those event tables we also glean information about the tool that created the data. Downstream we have one or more entries in the data_arrays table that carries the actual data extracted from the file. In the data_array table the descriptors field is an array that typically contains text with two elements in the 1-D array; name and units. In the values field we have a 1-D array containing all the data points from the file associated with the descriptor.

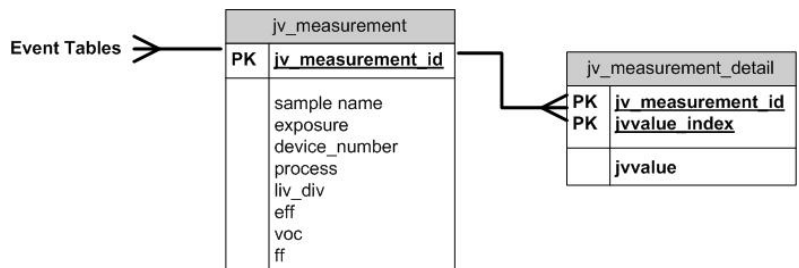

**Figure 7** –Example of an instrument semantic data model for a JV curve characterization instrument. This instrument produces x,y curves associated with devices on a PV sample.

The semantic database structures acknowledge known data semantics and relationships from a particular instrument. For example, an instrument may output standard x,y spectra related to a sample. The semantic model will take advantage of this known data relationship and tables will be created that relate x,y data to a specific sample. For instruments that are widely used for characterization, this is important as it allows for plotting and analysis tools to connect to the instruments data tables directly and any queries take advantage of database optimization automatically.

**Performance and Operations**

The LIMS implementation occurred over the last three years including a period of prototyping and testing. As expected we made adjustments to database table schemas, messaging and research access pipeline all focused on facilitating data quality and speed. After initiating basic raw data access, compatible instruments were then given translation configurations and all existing raw data files were post processed to extract the data and seed the initial database. From that point on all data harvesting and extraction is done as the instruments produce data files. No noticeable effects on performance and safety have been noted on any instrument and extracted data files are available to the researchers in typically under a minute, depending on current network traffic.

Since most extraction happens locally the data stream can be tapped during the archiving process and fed back into the system as a control process for the instrument. While we have yet to utilize this functionality, localized testing shows that we can get near real time ($< 1$ sec) response. This could be useful for in-situ measurements (e.g. ellipsometry) controlling some processing instruments despite no original integration between the two instruments. The system has been in full operation for the past two years with a duty cycle of greater than 95%.

The system has proven fault tolerant by ably handling the continuous ebb and flow of instruments moving from online to offline status. Additionally loss of network infrastructure or the server supporting the harvesting is quickly recovered once these systems come back on line. Monitoring software agilely finds differences in current and past harvesting activities and initiates the processes necessary to recover any missed data. Errors in data extraction can be repaired and the system triggered to correct any data errors. Over time, analysis of known data stream errors and methods of correction will be coded into future versions so it requires less human intervention.

**DISCUSSION**

There are many typical systems that preform rapid and routine observations for the researcher (e.g. transmission and reflectance measurements, JV characterizations, etc.). We have been able to identify those and have incorporated "data viewers" to enable the researchers to quickly assess data. Previously many of these systems required downloading the data to another computer and then processing it through conventional analysis means. By incorporating these quick views, researchers can save time and resources towards planning targeted work and not wasting resources following unneeded research paths [6]. We will soon have in place simple Boolean search tools to allow researchers to search and access files from across the spectrum of available instruments, projects and samples. Any of the basic search tools provided as part of the LIMS are of course granularity, but they can point out regions of interest with which data products can be identified, downloaded and processed through other "deep" data mining processes.

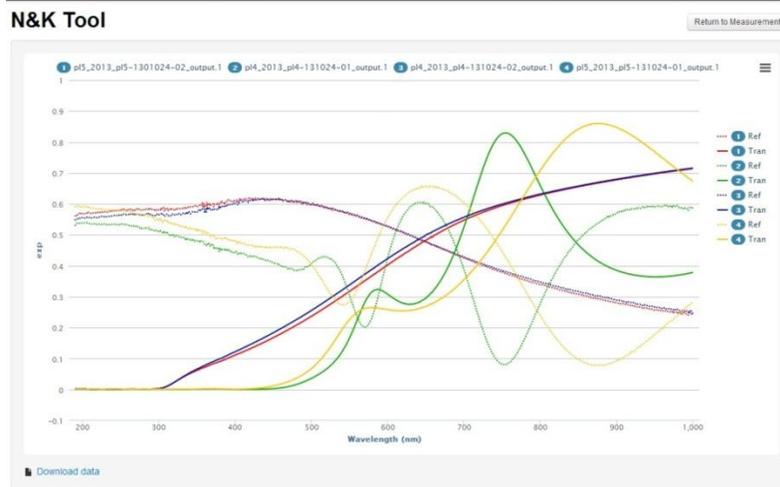

**Figure 8 -** Example of the Transmission and Reflectance viewer showing an interactive display space to visualize the data stored in the LIMS. Data from the instrumentation is harvested and stored in the LIMS database. The user can then access all of their data files for a particular instrument through a list-style interface. Selected files are then displayed through this viewing interface. Users can turn on and off traces, run multiple or singular graphs and overlays.

Our data stream archive was effectively used in a study in 2011 on multivariant AZO data. The data archive was mined to contribute to an understanding of various processing conditions and physical characteristics that were driving performance in TCOs [8].

**CONCLUSIONS**

Being able to handle large complex data sets is becoming a major focus in many scientific disciplines and at many intuitions. Advances in computer hardware has created a great flood of information that needs to kept securely, efficiently, and provide easily accessible data streams into advanced data mining tools. It's only through these data mining tools and their data streams that we will be able to effectively probe the ever growing parameter space supplied by instrumentation. A LIMS is the correct method for supporting this endeavor with its structured archiving and access capabilities. But development of a LIMS system to support a large integrated laboratory is not a task to take lightly. However the returns on a custom implementation of the software, in the case of the National Renewable Energy Lab solar program, far outweigh the disadvantages in not buying an off-the-shelf software solution. The implementation does take a large amount of time and resources but the software correctly fits the work flow and activities of the institution. This system has proven to be adaptable, agile, and fast in deploying new instruments onto the LIMS. In the future the experimental data stream from the NREL solar program will be able to merge with additional streams coming from other materials databases, theory driven modeling and simulation in an effort to provide a complete balanced strategy to supporting experimentation and theory [7].

**ACKNOWLEDGEMENTS**

This works was supported by the U.S. Department of Energy under Contract No. DE-AC36-08-G028308 with the National Renewable Energy Laboratory.